\documentclass[aps,prl,reprint,showpacs,preprintnumbers,amsmath,amssymb]{revtex4-1}
\pdfoutput=1
\usepackage{epsfig,graphicx,psfrag}
\usepackage{color}
\newcommand{\cO}{\mathcal{O}}
\newcommand{\no}{\nonumber \\}

\newcommand{\chpt}{$\chi$PT}

\newcommand{\mpi}{M_\pi}

\newcommand{\meta}{M_\eta}

\begin{document}


\title{The decay $\eta \to 3 \pi$: study of the Dalitz plot and extraction of the quark mass ratio $Q$}

\author{Gilberto Colangelo, Stefan Lanz, Heinrich Leutwyler}
\affiliation{Albert Einstein Center for fundamental physics, Institute for Theoretical Physics, University of 
Bern, Sidlerstr. 5, CH-3012 Bern, Switzerland}
\author{Emilie Passemar}\affiliation{Department of Physics, Indiana University, Bloomington, IN 47405, USA \\
Center for Exploration of Energy and Matter, Indiana University, Bloomington, IN 47403, USA \\
Theory Center, Thomas Jefferson National Accelerator Facility, Newport News, VA 23606, USA } 

\begin{abstract}
The $\eta\to 3\pi$ amplitude is sensitive to the quark mass difference
$m_u-m_d$ and offers a unique way to determine the quark mass ratio
$Q^2\equiv (m_s^2-m_{ud}^2)/(m_d^2-m_u^2)$ from experiment. We calculate
the amplitude dispersively and fit the KLOE data on the charged mode,
varying the subtraction constants in the range allowed by chiral
perturbation theory. The parameter-free predictions obtained for the
neutral Dalitz plot and the neutral-to-charged branching ratio are in
excellent agreement with experiment. Our representation of the transition
amplitude implies $Q = 22.0 \pm 0.7$. 
\end{abstract}

\pacs{}
\maketitle 

The decay $\eta \to 3 \pi$ is forbidden by isospin symmetry. Sutherland
\cite{Sutherland1966} showed that electromagnetic effects are only
subdominant with respect to the contribution coming from the up and down
quark mass difference $m_u-m_d$. A measurement of this decay can therefore
be used as a sensitive probe of the size of isospin breaking in the QCD
part of the Standard Model lagrangian. 

Chiral perturbation theory (\chpt)
offers a systematic method for the analysis of strong interaction processes 
at low energy. The chiral representation of the transition amplitude 
is known up to and including NNLO \cite{Osborn+1970,Gasser+1985a,Bijnens+2007}, 
but the expansion converges only very slowly. The reason is
well understood and has to do with rescattering effects in the final state
\cite{Khuri+1960,Roiesnel+1981}.  
As shown in \cite{Anisovich1995,*Anisovich:2013gha,Anisovich+1996,Kambor+1996}, 
these effects can reliably be
calculated with dispersion relations. In the meantime, the $\pi\pi$ phase shifts
have been determined to remarkable 
precision \cite{Colangelo2001,Descotes-Genon+2002,Kaminski:2006qe} and the
quality of the experimental information about $\eta\to3\pi$ is now much
better. This has triggered renewed interest in theoretical studies of
this decay~\cite{Gullstrom:2008sy,Schneider+2011,Kampf+2011,Lanz:2013ku, 
Guo:2015zqa,*Guo:2016wsi,Albaladejo+2015,Kolesar:2016jwe}.

The aim of the letter is to improve earlier dispersive treatments and to show
that this leads to a good understanding of these decays, 
in particular also to a better determination of $Q$.
Our analysis is based on three assumptions:

1. The dominant contribution to the transition amplitude is
proportional to $m_u-m_d$ with an isospin symmetric proportionality 
factor. We denote the dispersive representation of this contribution 
by $A_\mathrm{disp}(s,t,u)$ and normalize 
it to the mass difference between the charged and neutral kaons:
\begin{equation}
\label{eq:AtoM}
A_\mathrm{disp}(s, t, u) = -(M_{K^0}^2-M_{K^+}^2)_\mathrm{QCD} M(s, t, u)
\; .
\end{equation}
The function $M(s,t,u)$ concerns the isospin limit of QCD. We assume that
the remainder, which contains contributions due to the electromagnetic
interaction as well as terms of higher order in the isospin breaking
parameter $m_u-m_d$, can be accounted for with the one-loop
representation of \cite{Ditsche+2009}.

2. In the discontinuities of the amplitude, the $D$ and higher waves 
are strongly suppressed at low energies -- in the chiral expansion, 
they contribute only beyond NNLO.  Neglecting 
these contributions, the amplitude $M(s,t,u)$ can be decomposed 
into three functions of a single variable:
\begin{eqnarray}
\label{eq:RT}
M(s, t, u) &=& M_0(s) + (s-u) M_1(t) + (s-t) M_1(u) \no
&&+ M_2(t) + M_2(u) - \mbox{$\frac{2}{3}$} M_2(s) \; \; .
\end{eqnarray}
The three functions represent the $s$-channel isospin components 
of the amplitude ($I = 0,1,2$).  We expect representation
(\ref{eq:RT}) to constitute an excellent approximation to the exact
amplitude in the physical region of the decay. In this approximation, 
causality and unitarity lead to a set of dispersion relations, which determine the three
functions $M_I(s)$, in terms of the $S$- and $P$-wave phase shifts of $\pi
\pi$ scattering up to a set of subtraction constants.  

As is well known, the decomposition (\ref{eq:RT}) is unique only up to
polynomials. In particular, one may add an arbitrary cubic polynomial to
$M_2(s)$; the 
amplitude $M(s,t,u)$ stays the same provided
suitable cubic (quadratic) polynomials are added to $M_0(s)$
($M_1(s)$). Moreover, even if $M_2(s)$ is kept fixed, an ambiguity remains:
adding a constant to $M_1(s)$ leaves $M(s,t,u)$ unchanged, provided $M_0(s)$
is amended with a suitable term linear in $s$. This then exhausts the
degrees of freedom: the decomposition is unique up to a 5-parameter family
of polynomials. 

3. We fix the number of subtractions by imposing a condition on the
asymptotic behaviour: the function $M(s,t,u)$ is not allowed to grow more
rapidly than quadratically when the Mandelstam variables $s$, $t$, $u$
become large (notice that only two of the three variables are independent,
$s + t + u = M_\eta^2 + 3 M_\pi^2$).  

As demonstrated in \cite{Anisovich+1996}, the functions
$M_I(s)$ only have a right hand cut, with a discontinuity given by
\begin{equation} \label{eq:disc}
\mathrm{disc}M_I(s)= \left[M_I(s)+\hat{M}_I(s) \right] \sin \delta_I(s)
e^{-i\delta_I(s)}\;,
\end{equation}
where $\delta_I(s)$ is the phase of the lowest $\pi \pi$ partial wave of isospin $I$.
While the first term on the right hand side arises from collisions in the
$s$-channel, the second is generated by two-particle interactions in the
$t$- and $u$-channels and involves angular averages: detailed expression
can be found in \cite{Anisovich+1996}.  

It is advantageous to write dispersion relations not for the functions 
$M_I(s)$ but for $m_I(s) = M_I(s)/\Omega_I(s)$, where $\Omega_I(s)$
 is the Omn\`es factor belonging to $\delta_I(s)$:
\begin{equation}
\Omega_I(s)= \exp\left[ \frac{s}{\pi} \int_{4 M_\pi^2}^\infty ds'
  \frac{\delta_I(s')}{s'(s'-s-i\varepsilon)} \right],\; \; I=0,1,2 \;.
\end{equation}
This removes the first term on the right hand side of
 \eqref{eq:disc}: if the $t$- and $u$-channel discontinuities
 are dropped, $\mathrm{disc}\hspace{0.5mm}m_I(s)$  vanishes, so that  
 $m_I(s)$ represents a polynomial. More importantly: while 
 the dispersion relations for $M_I(s)$ admit nontrivial solutions even
 if the subtraction constants are set equal to zero, this does not happen
 with the dispersion relations for $m_I(s)$ -- in that case, the solution is uniquely determined
 by the subtraction constants. 
  
The only difference between the system of integral equations
that follows from the above assumptions and the one studied in
\cite{Anisovich+1996} is that
we are imposing a weaker asymptotic condition, that is, introduce
additional subtraction constants. The condition 3. fixes the amplitude up
to 11 subtraction constants. In view of the polynomial ambiguities, 5 of
these drop out in the sum, but the remaining 6 are of physical interest. 
Denoting these by $ \alpha_0$, $\beta_0$, $\gamma_0$, $\delta_0$, $\beta_1$
and $\gamma_1$, ($\delta_0$ and $\gamma_1$ are new compared to the
analysis in \cite{Anisovich+1996}) the integral equations take the form
\begin{equation}
\begin{split}
  M_0(s) = &\Omega_0(s) \left\{ \alpha_0 + \beta_0 s + \gamma_0 s^2 +
    \delta_0 s^3 + D_0(s) \right\} \; , \\
  M_1(s) = &\Omega_1(s) \left\{ \beta_1 s + \gamma_1 s^2 + D_1(s) \right\} \; , \\
  M_2(s) = &\Omega_2(s) D_2(s)  \; .
  \label{eq:dispRelFull}
\end{split}
\end{equation}
with
\begin{equation}
 \label{eq:DIdef}
D_I(s) = \frac{s^{n_I}}{\pi} \int_{4 \mpi^2}^\infty \frac{ds'}{s^{\prime n_I}}
	    \frac{ \sin \delta_I(s') \hat{M}_I(s') }{ |\Omega_I(s')| (s' -
              s -i \epsilon ) } \; ,
\end{equation}
where $n_0=n_2=n_1+1=2$.

As we are using many subtractions, the contributions arising from the high
energy part of the integrals in (\ref{eq:DIdef}) are not important.  
We could have made two additional subtractions
in the definition of the functions $D_I(s)$, so that their Taylor expansion
in powers of $s$ would only start at $\cO(s^4)$ for $I=0,2$ and at
$\cO(s^3)$ for $I=1$ -- this would merely change the significance of the
subtraction constants. The form chosen simplifies the comparison with
earlier work. For the same reason, the behaviour of the phase shifts at
high energies is irrelevant. We guide the phases to a multiple of $\pi$ at
$\sqrt{s}=1.7$ GeV. Since the integrands in (\ref{eq:DIdef}) are proportional to
$\sin(\delta_I)$, this implies that the integrals only extend over a
finite range -- with the number of subtractions we are using, convergence 
is not an issue.

If the subtraction constants as well as the phase
shifts are given, the integral equations impose a linear set of constraints
on the functions $M_I(s)$. Since the corresponding homogeneous system,
obtained by setting the subtraction constants equal to zero, does not admit
a nontrivial solution, the amplitude is determined uniquely: the general
solution of our equations represents a linear combination of the $3 \times
6$ basis functions $M_I^{\alpha_0}(s)$, $M_I^{\beta_0}(s)$, $\ldots$,
$M_I^{\gamma_1}(s)$, $I=0,1,2$: 
\begin{equation}
M_I(s)=\alpha_0 M_I^{\alpha_0}(s)+\beta_0 M_I^{\beta_0}(s)+\ldots+ \gamma_1
M_I^{\gamma_1}(s) \;.
\end{equation}
The basis functions can be determined iteratively -- the iteration converges 
in a few steps.  

While the effects due to $(m_u -m_d)^2$ are tiny, those from the
electromagnetic interaction are not negligible. In particular, the 
e.m.~self-energy of the charged pion generates a mass difference 
to the neutral pion which affects the phase space integrals quite 
significantly. 
 We estimate the isospin breaking effects with \chpt,
comparing the one-loop representation of the transition amplitude in
\cite{Ditsche+2009} (denoted by $M_{\mathrm{DKM}}$) with
the isospin limit thereof, i.e.~with the amplitude $M_{\mathrm{GL}}$ of 
\cite{Gasser+1985a}. For this purpose, we construct a purely kinematic map that 
takes the boundary of the isospin symmetric phase space into the boundary of 
the physical phase space for the charged mode. Applied to
$M_{\mathrm{GL}}$, this map yields an amplitude $\tilde{M}_{\mathrm{GL}}$
that lives on physical phase space and has the branch points 
of the two-pion cuts at the proper place. The ratio $K\equiv
M_{\mathrm{DKM}}/\tilde{M}_{\mathrm{GL}}$  
is approximately constant over the entire Dalitz plot: in the charged
decay mode, this ratio only varies in the range $1.031 < |K|^2 < 1.078$.
In the neutral channel, the branch cuts from the transition 
$\pi^0\pi^0\to\pi^+\pi^-\to \pi^0\pi^0$ run through the physical region. 
Our method accounts for these only to NLO, via the factor $K$, but 
the narrow range $0.972<|K|^2< 0.978$ shows that their contributions
are numerically very small. In both decay modes, the normalized Dalitz  
plot distribution of $\tilde{M}_{\mathrm{GL}}$ is remarkably close to the 
one of the full one-loop representation. 

In this sense, the distortion of phase space generated by the self-energy 
of the charged pion dominates the isospin-breaking effects in the Dalitz 
plot distribution. We denote the amplitude obtained from our  isospin 
symmetric dispersive representation $A_\mathrm{disp}$ with the map introduced 
above by $\tilde{A}$ and approximate the physical amplitude with 
$A_{\mathrm{phys}}= K\tilde{A}$.  As discussed below, the prediction 
obtained for the branching ratio of the two modes provides a stringent 
test of this approximate formula: the factor $|K|^2$ barely affects 
the Dalitz plot distribution because it is nearly constant, but it differs
from unity and therefore affects the rate.  
Details will be given in \cite{CLLP}.  

The experimental results on the Dalitz plot distribution do not suffice to
determine all subtraction constants. In particular, the overall
normalization of the amplitude is not constrained by these. We use the
one-loop representation of \chpt~ to constrain the admissible range of the
subtraction constants. To do this we consider the Taylor coefficients of
the functions $M_0(s)$, $M_1(s)$ and  $M_2(s)$: 
\begin{equation}
M_I(s) = A_I+ B_I s+ C_I s^2+ D_I s^3+ \ldots 
\end{equation}
These coefficients also depend on the choice made in the decomposition
 (\ref{eq:RT}) of the one-loop representation, but the combinations 
\begin{equation}
\begin{split}
H_0 =& A_0 + \frac{4}{3} A_2 + s_0 \left(B_0 + \frac{4}{3} B_2\right) \\
H_1 =& A_1 + \frac{1}{9}\left(3 B_0 -5 B_2\right)- 3 C_2 s_0 \\
H_2 =& C_0 + \frac{4}{3} C_2 , \qquad H_3 = B_1 + C_2 \\
H_4 =& D_0 + \frac{4}{3} D_2 , \qquad H_5 = C_1 - 3 D_2
\end{split}
\end{equation}
are independent thereof ($s_0$ defines the center of the Dalitz plot: $s_0
= \frac{1}{3} \meta^2+\mpi^2$). We use the constant $H_0$ to parameterize the 
normalization of the amplitude and describe the relative size
of the subtraction constants  by means of the variables $ h_I = H_I/H_0$. 
Specifying the 6 threshold coefficients $H_0,h_1,\ldots\,,h_5$ is
equivalent to specifying the 6 subtraction constants $\alpha_0$, $\beta_0$,
$\ldots$, $\gamma_1$.  

At leading order of the chiral expansion, only $H_0^\mathrm{LO} = 1$ and
 $h_1^\mathrm{LO}= 1/(\meta^2 - \mpi^2)=3.56$ are different from zero 
 (throughout, dimensionful quantities are given in GeV units). 
The NLO representation yields corrections for these two coefficients as
well as the leading terms in the chiral expansion of $h_2$ and $h_3$. The
one-loop formulae can be expressed in terms of the masses, the 
decay constants $F_\pi,F_K$ and the low energy constant $L_3$, which
only contributes to $H_3$.  We are using the recently improved determination 
$L_3=-2.63(46)\cdot 10^{-3}$ of \cite{Colangelo:2015kha}, so that the 
one-loop representation does not contain any unknowns.

Experience with $\chi$PT indicates that, unless the quantity of interest 
contains strong infrared singularities, subsequent terms in the chiral 
perturbation series based on $SU(3)\times SU(3)$ are smaller by a
factor of $20-30\%$. The values  $H_0^\mathrm{NLO} = 1.176, 
h_1^\mathrm{NLO}=4.52$ confirm this rule: while in the case of $H_0$, the 
correction is below 20\%, the one in $h_1$ is relatively large (27\%), 
because this quantity does contain a strong infrared singularity: $h_1$ 
diverges in the limit $\mpi \to 0$, in proportion to $1/\mpi^2$. In fact, 
the singular contribution fully dominates the correction.  
We conclude that it is meaningful to truncate the chiral expansion of the 
Taylor coefficients at NLO. The invariant $X$ is approximated with
the one-loop result $X^\mathrm{NLO}$ and the uncertainties
from the omitted higher orders are estimated at
$0.3\,|X^\mathrm{NLO}-X^\mathrm{LO}|$. This is on the conservative
side of the rule mentioned above and yields a theoretical estimate 
for four of the six coefficients:
$H_0=1.176(53)$, $h_1= 4.52(29)$, $h_2= 16.4(4.9)$, $h_3=6.3(1.9)$
(the estimate used for $h_3$ in particular also covers the comparatively 
small uncertainty in the value of $L_3$). The remaining two are beyond 
reach of the one-loop representation -- we treat $h_4$ and $h_5$ as 
free parameters.

The observed Dalitz plot distribution offers a good check of
these estimates: dropping the subtraction constants $\delta_0,
\gamma_1$ and ignoring $\chi$PT altogether, we obtain a three-parameter 
fit to the KLOE Dalitz plot with $\chi^2_\mathrm{exp}=
 385$ for 371 data points. For all three coefficients $h_1,h_2,h_3$,
 the fit yields a value in the range estimated above on the basis of $\chi$PT. 
 Moreover, along the line $s=u$, the resulting representation for the real part of the 
 amplitude exhibits a zero at $s_A^\mathrm{fit}=1.43 M_\pi^2$: 
 the observed Dalitz plot distribution implies the presence 
 of an Adler zero, as required by a venerable $SU(2)\times
SU(2)$ low-energy theorem \cite{Cronin1967} (at leading order of the chiral expansion, 
the zero  sits at  $s_A^\mathrm{LO}=\frac{4}{3}M_\pi^2$). 
 
The three assumptions formulated above do not imply that the subtraction
constants are real. In fact, beyond NLO of the chiral expansion, the
subtraction constants get an imaginary part which can be estimated with the
explicit expressions obtained from the two-loop representation: they do not
contain any unknown LECs, and none of the $\cO(p^6)$ ones. For simplicity,
we take $\alpha_0, \beta_0,\ldots,\gamma_1$ to be real. 
The small changes occurring if the imaginary parts of the subtraction
constants are instead taken from the two-loop representation barely affect
our results.

In our analysis, the recent KLOE data \cite{KLOE:2016qvh} play the central 
role. In this experiment, the Dalitz plot distribution of the decay 
$\eta \to \pi^+ \pi^- \pi^0$ is determined to high accuracy, bin-by-bin. 
In the following we restrict ourselves to an analysis of these data. 
The results of earlier experiments \cite{Ambrosino:2008ht,Adlarson:2014aks,
Ablikim:2015cmz} can readily be included, but do not have a significant 
effect on our results \cite{CLLP}.  

We minimize the sum of two discrepancy
functions: while $\chi^2_{\mathrm{exp}}$  measures the difference between 
the calculated and measured Dalitz plot distributions at the 371 data points of
KLOE~\cite{KLOE:2016qvh}, $\chi^2_{\mathrm{th}}$ represents the sum of the square
of the differences between the values of $h_1$, $h_2$ and $h_3$ used in the
fit and the central theoretical estimates, divided by the uncertainties
attached to these. The minimum $\chi^2=\chi^2_{\mathrm{exp}}+\
\chi^2_{\mathrm{th}}$ we obtain for the 371 data points is
equal to $\chi_\mathrm{exp}^2=380.2$, at the parameter values (the
subtraction constants are univocally fixed by these):
\begin{equation}\label{eq:results h_i}
\begin{split}
&\mbox{Re}h_1=4.49(14),\,\mbox{Re}h_2=21.2(4.3),\,
\mbox{Re}h_3=7.1(1.7),\,\\
&\mbox{Re}h_4=76.4(3.4),\,\mbox{Re}h_5=47.3(5.8) \; .
\end{split}
\end{equation}
The quoted errors are based on the Gaussian approximation.
The noise in the input used for the phase shifts generates 
an additional contribution. To estimate it, we have varied the 
Roy solutions of \cite{Colangelo2001}, not only below
800 MeV where the uncertainties are small, but also at higher 
energies where dispersion theory does not provide strong constraints. 
The resulting fluctuations in the Taylor invariants are small compared
to the Gaussian errors obtained with the central input for the phase shifts  
-- the errors quoted in \eqref{eq:results h_i} include these uncertainties.

Our dispersive representation passes a crucial test: the real part of the 
transition amplitude does have a zero, remarkably close to the place 
where it was predicted on the basis of current algebra: $s_A=1.34(10) M_\pi^2$. 
The theoretical constraints play a significant role here: if
$\chi^2_\mathrm{th}$ is dropped, the quality of the fit naturally improves 
(the discrepancy with the KLOE data drops from 380 to 370), but outside 
the physical region, the parameterization then goes astray. In particular,
the Adler zero gets lost: with 5 free parameters in the representation
of the Dalitz plot distribution, the data do not provide enough information 
to control the extrapolation to the Adler zero.   

The solution \eqref{eq:results h_i} yields a parameter free
prediction for the Dalitz plot of the neutral channel.
\begin{figure}\includegraphics[width=8cm]{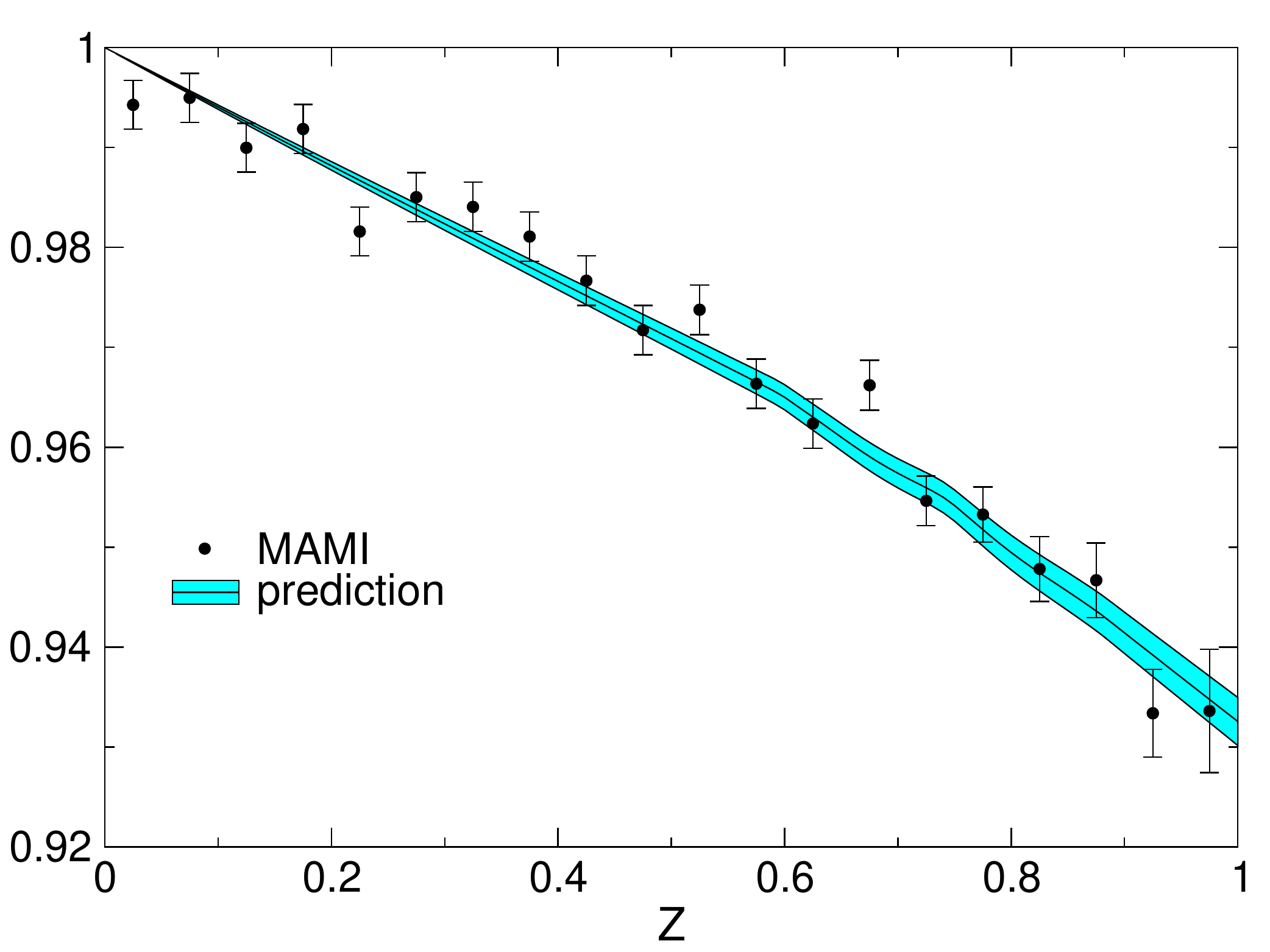}
\caption{Prediction obtained from 
the KLOE measurements of $\eta\to\pi^+\pi^-\pi^0$ compared with the MAMI results 
for $\eta\to 3\pi^0$ ($Z\equiv X^2+Y^2$ represents the square of the distance
from the center of the Dalitz plot).}\end{figure}
The figure shows that the resulting $Z$-distribution is in excellent agreement with the 
MAMI data  \cite{Prakhov+2009}. Quantitatively, the comparison 
yields $\chi^2=22.5$ for 20 data points (no free parameters).

This solves a long-standing puzzle:  \chpt~predicts 
the slope $\alpha$ of the Z-distribution to  be positive at one loop, while the
measured slope is negative. The problem arises 
because $\alpha$ is tiny -- estimating the uncertainties inherent in the one-loop 
representation with the rule given above, we find that the error in
$\alpha$ is so large that not even 
the sign can reliably be determined. The situation
does not improve at NNLO \cite{Bijnens+2007}. Only with dispersion theory
is one able to reach the necessary precision and to reliably predict the
slope. 

At the precision at which the slope
is quoted by the PDG,  $\alpha_\mathrm{PDG}=-0.0315(15)$ \cite{Agashe:2014kda},
the definition of  $\alpha$ matters, because the $Z$-distribution 
is well described by the linear formula $1+2\hspace{0.3mm}\alpha\hspace{0.1mm} Z$ only 
at small values of $Z$. For the slope at $Z=0$, we find
$\alpha=-0.0302(11)$, while a linear fit on the intervals $0<Z<0.5$ and
$0<Z<1$ yields the slightly different values 
 $\alpha=-0.0293(11)$ and  $\alpha=-0.0313(11)$, respectively.

 The decay rates of the processes $\eta \to \pi^+ \pi^- \pi^0$ and $\eta
 \to 3 \pi^0$ are given by an integral over the square of the corresponding
 amplitudes and hence by a quadratic form in the subtraction constants. 
 For the individual rates, $H_0$ is also
 needed -- and will be discussed below -- but in the branching ratio $B
 =\Gamma(\eta \to 3 \pi^0)/\Gamma(\eta \to \pi^+ \pi^- \pi^0) $ the
 normalization drops out. The uncertainties in the dispersive representation
 also cancel almost completely. The error in our result, $B=1.44(4)$, is 
 dominated by the uncertainties in the one-loop approximation used for
 isospin breaking.  The comparison with 
 the experimental values given by the
 Particle Data Group, $B = 1.426(26)$ [`our fit'], $B = 1.48(5)$ [`our
 average'] shows that the value predicted for the decay rate of the
 neutral mode (on the basis of Dalitz plot distribution and decay rate of
 the charged mode) agrees with experiment. This provides a very strong test of
 the approximations used to account for isospin breaking.  
  
 In contrast to the Dalitz plot distributions and the branching ratio, the
 individual rates do depend on the normalization of the amplitude, which we
 specify in terms of $(M_{K^0}^2-M_{K^+}^2)_\mathrm{QCD}$ and $H_0$. With
 the theoretical estimate for $H_0$ given above, the experimental values of
 the rates $\Gamma(\eta \to \pi^+ \pi^- \pi^0)=300(12)$ eV and $\Gamma(\eta
 \to 3 \pi^0)= 428(17)$ eV \cite{Agashe:2014kda} yield two separate
 determinations of the kaon mass difference in QCD. Since our prediction
 for the branching ratio agrees with experiment, the two results are nearly
 the same, but they are statistically independent only with regard to the
 uncertainties in the rates, which are responsible for only a small
 fraction of the error.  Combining the two, we can determine the mass
 difference to an accuracy of 6\%:
\begin{equation}\label{eq:DMK}
(M_{K^0}^2-M_{K^+}^2)_\mathrm{QCD}=  6.27(38) 10^{-3}\,\mbox{GeV}^2\,.
\end{equation} 
The comparison with the observed mass difference implies 
$(M_{K^0}^2-M_{K^+}^2)_\mathrm{QED}=-2.38(38) 10^{-3}\,\mbox{GeV}^2$.
The corresponding result $\epsilon=0.9(3)$ for the parameter used to measure
the violation of the Dashen-theorem \cite{Aoki:2016frl}, agrees with recent lattice results 
\cite{Fodor:2016bgu,Basak:2016jnn} which also indicate that this theorem
picks up large corrections from higher orders. Indeed, the direct determination 
of $\epsilon$ based on an evaluation of the kaon mass difference with the 
e.m.~effective Lagrangian encounters unusually strong 
logarithmic infrared singularities, which generate large nonleading terms
in the chiral perturbation series  \cite{Langacker:1974nm}. We emphasize 
that our determination of $\epsilon$ does not face this problem.

Finally, we invoke the low energy theorem that relates the kaon 
mass difference to the quark mass ratio $Q$ \cite{Gasser+1985a}:
\begin{equation}\label{eq:DMQ}
(M_{K^0}^2-M_{K^+}^2)_\mathrm{QCD}=\frac{M_K^2(M_K^2-\mpi^2)}{Q^2
  \mpi^2}  \,,
\end{equation}
($M_K$ and $\mpi$ stand for the QCD masses in the limit $m_u = m_d$).
Since the relation holds up to corrections of NNLO, our analysis goes through 
equally well if the quantity $(M_{K^0}^2-M_{K^+}^2)_\mathrm{QCD}$ is replaced by 
the right hand side of \eqref{eq:DMQ}. This leads to 
\begin{equation}\label{eq:Q} Q = 22.0(7)\,,
\end{equation} 
in good agreement with the values obtained on the lattice~\cite{Aoki:2016frl}. 
Using the remarkably precise result for the ratio $m_s/m_{ud}= 27.30(34)$ 
quoted in the same reference, we can finally also
determine the relative size of the two  lightest quark masses: $m_u/m_d=0.44(3)$.
The theoretical estimates for $H_0,h_1,h_2,h_3$ and the experimental
uncertainties in Dalitz plot distribution and rate contribute about
equally to the quoted error in this determination of the isospin breaking
quark mass ratios, while the uncertainties due to the noise in the $\pi\pi$ 
phase shifts are negligibly small. We defer a 
detailed discussion and a comparison with related work 
\cite{Bijnens+2007,Gullstrom:2008sy,Schneider+2011,Kampf+2011,
Guo:2015zqa,Albaladejo+2015,Kolesar:2016jwe,Guo:2016wsi} to a 
forthcoming publication \cite{CLLP}. 
 
\begin{acknowledgments}
We thank P.~Adlarson, J.~Bijnens, L.~Caldeira
Balkest\r{a}hl, I.~Danilkin, J.~Gasser, K.~Kampf, B.~Kubis, A.~Kup\'s\'c,
S.~Prakhov, A.~Rusetsky and P.~Stoffer for useful information.  This work
is supported in part by Schweizerischer Nationalfonds and the
U.S. Department of Energy (contract DE-AC05-06OR23177).
\end{acknowledgments}

\bibliography{biblio}

\end{document}